\documentclass{PoS}

\usepackage{url}

\title{Eventdisplay: An Analysis and Reconstruction Package for Ground-based Gamma-ray Astronomy}

\ShortTitle{Eventdisplay: An Analysis Package for Gamma-ray Astronomy}

\author{\speaker{Gernot Maier}$^{1}$ and Jamie Holder$^{2}$\\
        $^{1}$ DESY, Platanenallee 6, D-15738 Zeuthen, Germany\\
        $^{2}$ Department of Physics and Astronomy and the Bartol Research Institute, University of Delaware, Newark, DE 19716, USA \\
        E-mail: \email{gernot.maier@desy.de}}

\abstract{
Eventdisplay is a software package for the analysis and reconstruction of data and Monte Carlo events from ground-based gamma-ray observatories such as VERITAS and CTA. It was originally developed as a display tool for data from the VERITAS prototype telescope, but evolved into a full analysis package with routines for calibration, FADC trace integration, image and stereo parameter analysis, response function calculation, and high-level analysis steps. Eventdisplay makes use of an  image parameter analysis combined with gamma-hadron separation methods based on multivariate algorithms. An overview of the reconstruction methods and some selected results are presented in this contribution.}

\FullConference{35th International Cosmic Ray Conference\\
		 12-20 July, 2017\\
		 Bexco, Busan, Korea}

\begin{document}

\section{Introduction}

Very-high-energy gamma rays, i.e.~photons with energies in the range of 50 GeV to hundreds of TeV, are measured routinely with ground-based imaging atmospheric Cherenkov telescopes (IACTs)  from a large variety of  galactic and extragalactic objects.
These measurements allow to trace and and identify the acceleration processes of high-energy particles at, or close to, their acceleration sites.

The IACT technique was pioneered by the Whipple collaboration using a 10-m reflector located on Mt. Hopkins, Arizona \cite{Weekes-1989}. 
The Whipple 10-m telescope was in operation from 1968 to 2011 and sensitive to gamma rays in the energy range from 200 GeV to 20 TeV. 
The first very-high-energy gamma ray sources were detected with this instrument: the Crab Nebula \cite{Weekes-1989} and Mrk 421 \cite{Punch-1992}.
New hardware such as the deployment of fast electronics and sophisticated trigger schemes contributed to the success of Whipple.
However, the breakthrough in the field which lead to the detection of the first gamma-ray source was the development of new analysis methods using Monte Carlo simulations by Hillas \cite{Hillas-1985}.
The field has expanded hugely since then, with many more detailed and sophisticated measurements having been made.
Today, in 2017, three major telescope array systems are in operation: 
H.E.S.S.\footnote{http://www.mpi-hd.mpg.de/hfm/HESS/},
MAGIC\footnote{http://magic.mppmu.mpg.de/}, and 
VERITAS\footnote{https://veritas.sao.arizona.edu/}.

IACTs collect the Cherenkov light emitted when relativistic particles (mainly electrons and positrons) from air showers pass through the atmosphere. 
The Cherenkov photons are produced along the shower axis with an emission maximum at about 10 km above ground; they form a short, nanoseconds long, bluish (300-550 nm) light flash. 
The emission angle changes with altitude from about 0.1 deg at 30-km height a.s.l. to about 1.3 deg at sea level. 
This results in an approximately flat lateral distribution of Cherenkov photons on the ground with a radius of roughly 140 m. 
Placing a single telescope anywhere inside the light pool results in a large sensitive detection area of about $10^5$~m$^2$.
Each IACT measures a projection of the longitudinal development of the air shower in its camera. 
Using  different reconstruction techniques, the most important characteristics of the astrophysical photons are derived:  the arrival direction, from the orientation of the shower images, and the primary energy, from the total signal size.
Most events measured are initiated by charged cosmic rays and not by gamma rays. 
Cosmic-ray air showers are in general much more irregular with significantly more energy transferred to larger lateral distances from the shower axis. 
Most of these background events can therefore be eliminated by applying analysis cuts on the shape of their images in the camera. 
Modern ground-based gamma-ray instruments consist of arrays of several telescopes, which improves the above processes significantly. 
VERITAS consists of four IACTs and the Cherenkov Telescope Array (CTA) will consist of two arrays with 29 and 99 telescopes of different sizes equipped with different types of cameras and readout electronics.

These proceedings describe the Eventdisplay software, a package developed to display, calibrate, reconstruct, and analyse data and Monte Carlo events from ground-based gamma-ray observatories such as VERITAS and CTA\footnote{https://www.cta-observatory.org} using some of the techniques originially developed by Hillas.

\section{Eventdisplay}

The Eventdisplay software is one of the main software packages developed and used  for
the analysis of data and simulations in the VERITAS collaboration \cite{Daniel-2007}.
It was originally designed as an event display for the VERITAS prototype data, but evolved into
a full package including all relevant steps for reconstruction and analysis.
The software has been utilised for the analysis of Monte Carlo simulations of the response of the 36-telescope concept of AGIS \cite{Maier-2009} and is now used for the analysis of CTA arrays with a large number of telescopes. 
Eventdisplay is written in C++ and  makes use of the data analysis routines provided by the ROOT scientific software framework\footnote{https://root.cern.ch/}.
The Eventdisplay software is free to use for anybody and available on request from the author.

The performance of  Eventdisplay has been extensively tested on VERITAS data and simulations (see e.g.~\cite{Maier-2007}).
The sensitivity achieved is sufficient to detect sources with a flux of 1\% of the Crab Nebula in less than 25 hr of observations with the VERITAS 
observatory\footnote{see http://veritas.sao.arizona.edu/about-veritas-mainmenu-81/veritas-specifications-mainmenu-111}.
The VERITAS data evolved significantly over time with: changes in data formats, detector upgrades (relocation of one telescope; upgrade of the telescope cameras), and in the parameters of the data acquisition and calibration. 
This led to a flexibility in the Eventdisplay package which made it relatively easy to adapt the software for new reconstruction methods and for next-generation  instruments with a larger number of telescopes and a variety of telescope designs, such as CTA.
The main steps of the analysis are described in the following. 

\begin{figure}
\centering\includegraphics[width=0.95\linewidth]{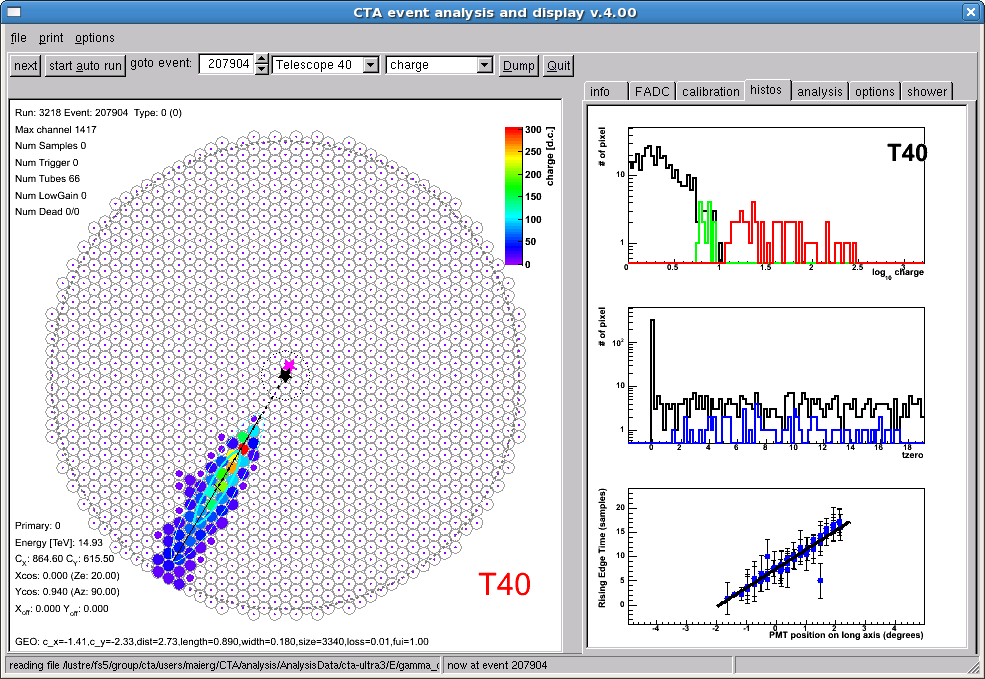}
\centering\includegraphics[width=0.95\linewidth]{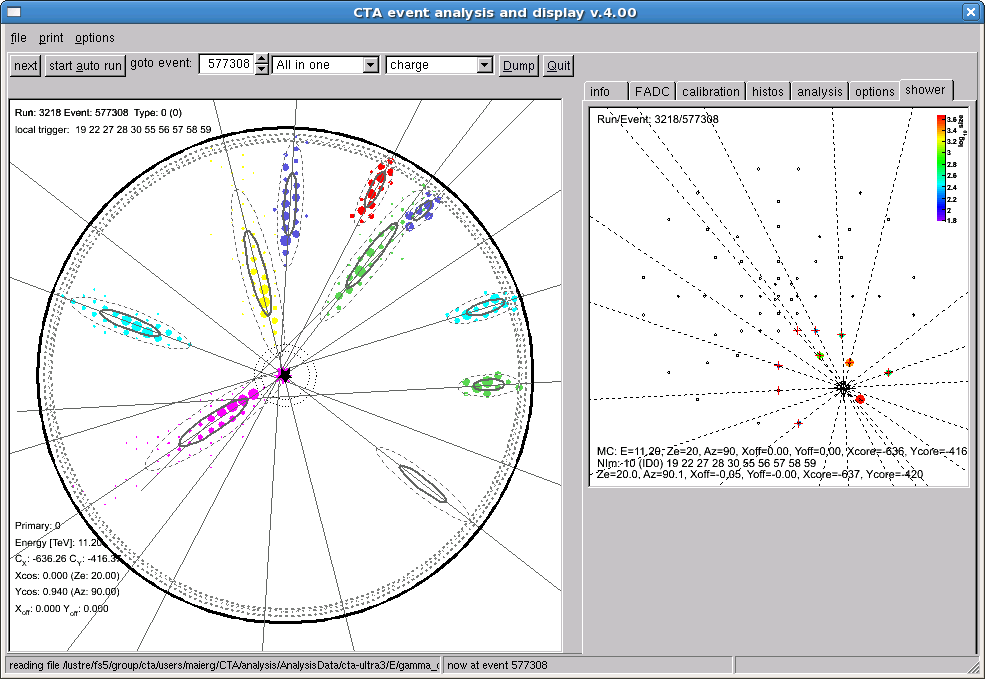}
\caption{\label{display}
Camera and event display for a randomly selected high-energy event from simulations of a CTA medium-sized telescope.
The panel to the top right shows (from top to bottom) the distribution of integrated charges (black line: all; red: image pixel; green: border pixel), the distribution of pulse arrival times (50\% rise time) for image and border pixels, and the time gradient along the long axis of the image used for the `double-pass` trace integration. The bottom figure shows the camera images (left) and the distribution of telescopes on the ground (right) from the same simulated event with primary energy of 11.2 TeV.}
\end{figure}

\subsection{Calibration}

The calibration routines in Eventdisplay include tools to analyse  events from the uniform illumination of the PMT cameras by LEDs or laser flashes. This analysis calculates relative gains of the pixels, timing differences due to path-length differences in the cabling of each channel, and the relative calibration of the high- and low-gain readout chain.
The centre component of the VERITAS data acquisition is a 500 MHz FADC system. Typically, a FADC window of 16-30 sample lengths is readout and stored on disk.  The characteristics of the planned readout systems for the CTA cameras vary significantly:  from 250 MHz to 1 GHz FADC systems to customised peak-detector electronics \cite{Ong-2017}.
Eventdisplay can handle these differences and provide the appropriate calibration:
the estimation of the voltage offset (pedestals) in each FADC trace and the noise levels due to the night sky background and electronics can be calculated either from artificially triggered `pedestal events` or from the triggered shower data.
These pedestal levels are calculated in short time bins (typically 3 min long), to take into account the variations of the background light level during  observing runs.
IACTs detect also the Cherenkov light emitted due to individual muons, which can be used for the calibration of the total optical throughput of the telescopes.
Several routines to reconstruct muons are part of Eventdisplay, including a muon identification method using Hough transforms \cite{Tyler-2013}.

\subsection{Signal extraction}

The signal (integrated charge) per pixel and the pulse arrival time is extracted from the integration of the relevant parts of the  FADC trace.
Several methods for this are implemented in Eventdisplay, including the integration using a fixed position of the integration window, a sliding window method searching for the maximum signal along the trace, and a two-pass method which uses the time gradient along the measured image to find the optimal position for the integration window \cite{Holder-2006}.
An example for such a time gradient across the long axis of an image can be seen in Figure 1 (top right panel).
The length of the integration window is configurable and should be chosen according to the typical width of the pulses provided by the readout system.

\subsection{Image analysis}

Images are cleaned to identify the signal from the Cherenkov light emitted in the air shower and to remove that due to sky brightness fluctuations.
Several different cleaning algorithms are available in Eventdisplay, including the traditional cleaning consisting of a two-level filter with user-defined thresholds $q_1$ and $q_2$ ($< q_1$) \cite{Hillas-1998}.
This filter removes all pixels with an integrated charge smaller than $q_1$ (image pixels) and any pixels that are adjacent to the remaining pixels and have signals smaller than $q_2$ (border pixels). 
Besides the described cleaning with fixed image and border thresholds, variable cleaning levels depending on the average noise level per pixel can be applied as an cleaning method.
This is the default cleaning method applied in the analysis of VERITAS data and proofed to be robust against the illumination of sections of the  camera by bright stars, and for observations taken under elevated background light conditions during bright moon phases. 
Several variations of these cleaning schemes are available (e.g.~removal of small clusters of image/border pixels \cite{Bond-2003}, or the application of additional signal arrival time constraints).
Additionally, the optimal next-neighbour cleaning method \cite{Shayduk-2013},  applying dynamical cuts in charge and time space, is implemented in Eventdisplay.
The cleaning levels in this method are derived from pixels rates determined from night-sky background events.
This cleaning method delivers a remarkable gain in low-energy events and a much lower analysis threshold in energy (see  \cite{Shayduk-2013}).

The shower image is then parametrised with a second moment analysis \cite{Hillas-1985}. 
A  log-likelihood fitting algorithm is applied to recover partially contained images at the edge of the camera.
The underlying assumption of the fitting method is that the image of a gamma-ray shower can be described by a two-dimensional normal distribution. 
This extrapolation leads to an increase in sensitivity for showers with large energies or large directional offsets relative to the camera center.
This fitting method, applied not only to the events at the edge of the camera, allows additionally to minimise the influence of dead or suppressed channels on the estimation of image parameters. 

\subsection{Direction, shower core, and energy reconstruction}

The direction of origin of the very-high-energy gamma ray on the sky and the impact parameter of the shower core on the ground are reconstructed using stereoscopic techniques \cite{Hofmann-1999, Krawczynski-2006}.
The shower direction is calculated from the mean intersection points between all possible pairs of telescope images. Image pairs are weighted by image size, elongation, and their relative angle to each other (giving less weight to image pairs with almost parallel orientation).
Alternatively, a method based on the calculation of the displacement (algorithm 3 in \cite{Hofmann-1999}) of the shower direction from the centroid of each image is available.
The displacement is calculated by regression boosted decision trees \cite{TMVA}, trained with simulated gamma-ray events using mainly the width, length, and image size image parameters.
This algorithm performs also very well for events with mostly parallel images, e.g.~at large zenith angle observations or for large off-axis events.

The energy of each gamma-ray is estimated from Monte Carlo simulations. 
The calculation uses lookup tables \cite{Krawczynski-2006} or regression boosted decision trees and determines the energy of an event as a function of impact parameter, integrated charge per image, level of night-sky background, azimuth and zenith angle, and array configuration.

\subsection{Gamma-selection techniques}

\begin{figure}
\centering\includegraphics[width=0.98\linewidth]{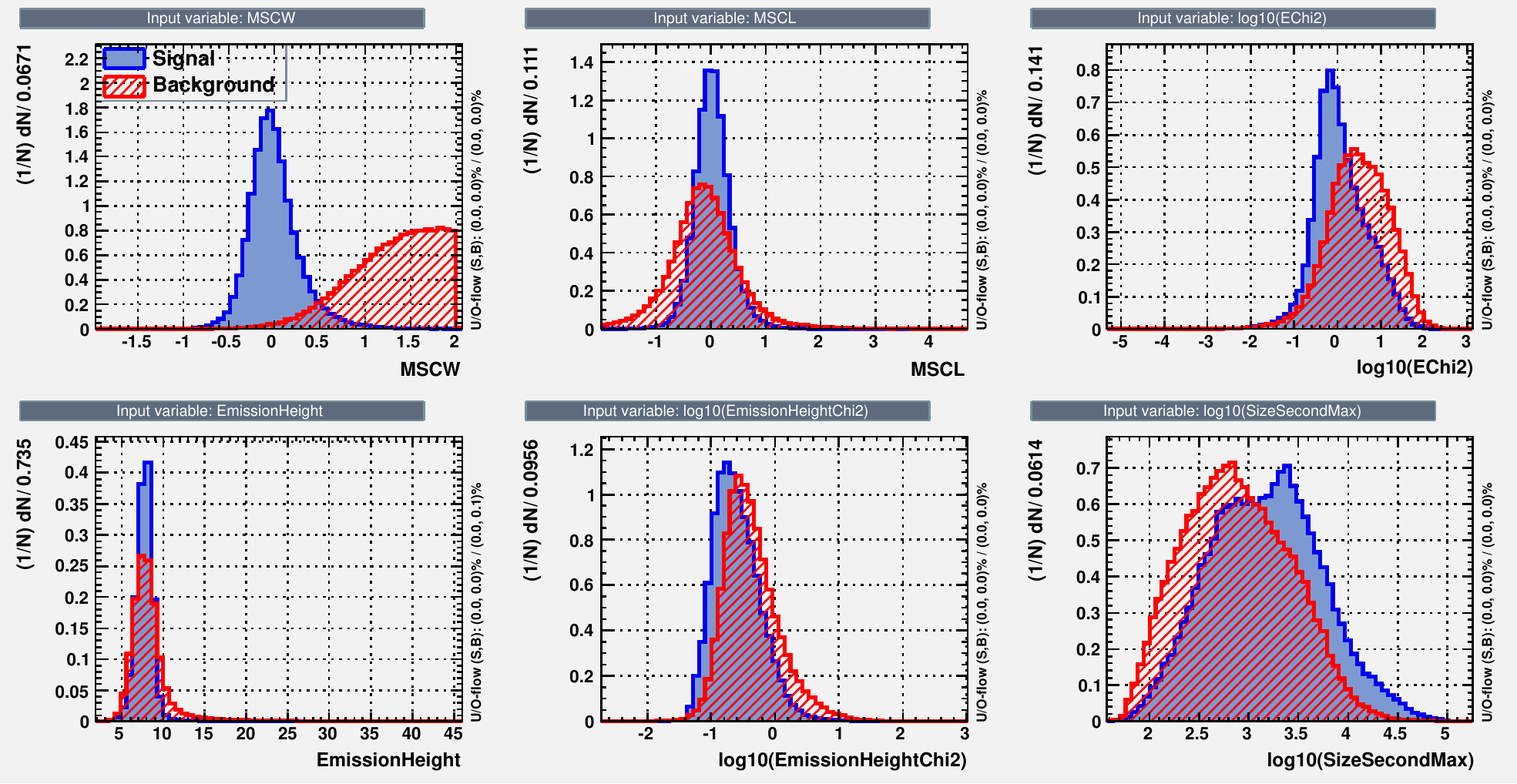}
\caption{\label{CutVar:highE}
Distribution of cut variables  for events with high energies ($\approx 10$ TeV). Signal events are simulated gamma rays, background events consist of proton events.
See text for a description of the parameters.
}
\end{figure}

The majority of the far more numerous background events due to cosmic rays are rejected by comparing the shape of the event images in each telescope with the expected shapes of  gamma-ray showers determined with MC simulations.
Several variables are used for gamma-hadron separation (all are explained below): mean-scaled width and length, shower direction, spread of the energy reconstruction, and height of the maximum Cherenkov emission.

Mean-scaled width and mean-scaled length parameters \cite{Krawczynski-2006} are calculated in Eventdisplay using lookup tables based on MC simulations of gamma rays.
The height of the maximum Cherenkov emission is determined by triangulation using the centroid position of the images and the telescope positions.
Energies are reconstructed per telescope, as described in the previous section.
For gamma rays, the spread in reconstructed energy between the telescopes is much lower than for hadrons.
The reason for this is the irregularity of the hadronic showers, which leads to larger fluctuations in the measured Cherenkov light at the location of each telescope.
Figure \ref{CutVar:highE} shows the distribution of these variables for gamma-ray and proton events as determined from MC simulations for events with reconstructed energies of about 10 TeV (the shape and mean values change with energy).

There are a variety of options for the gamma-ray hadron separation routine in Eventdisplay, ranging from box cuts using single-telescope or stereo parameters to energy-dependent cuts based on multivariate methods:
Boosted decision trees (BDT) as implemented in the TMVA package \cite{TMVA} show the best performance of the used gamma-hadron separation methods. 
The BDT can be trained and applied in several energy and off-axis bins; see \cite{Krause-2017} for an extensive description of the BDT analysis and performance.

\subsection{Science analysis: sky maps, spectral reconstruction, and light curves}

Eventdisplay contains several routines for the steps from event data (described in the previous sections) to the typical science results consisting of sky maps, energy spectra, and light curves.
The background at each position in the sky can be determined using the well-known reflected region or ring-background models. 
The algorithms allow the exclusion of regions of bright stars or other gamma-ray sources in the field of view.
Energy spectra and fluxes can be determined in Eventdisplay with the correction method (using the effective collection area as a function of reconstruction energy \cite{Mohanty-1998}) or with forward-folding methods \cite{Piron-2001}.

A recent addition is the possibility of writing the event data and  instrument response functions in FITS format following the open gamma-ray data format\footnote{https://gamma-astro-data-formats.readthedocs.io/en/latest/}.
This allows to use other community-developed science tools (e.g.~ctools \cite{ctools}) and apply, for example, likelihood procedures for the detailed modelling of the observed signal and background event distribution.

\section{Conclusions}

Eventdisplay is a versatile toolbox for reconstruction and data analysis for ground-based gamma-ray astronomy.
The software package has been used for many years for the analysis of VERITAS data and CTA MC events.
It can be relatively easily extended and used to test new reconstruction methods.
The software is freely available and can, in principle, be used for the analysis of data from any IACT instrument.

\acknowledgments

Eventdisplay was developed mostly for the analysis of VERITAS data. 
We gratefully acknowledge the support and valuable input through discussions, bug fixes, and substantial code contributions from many colleagues in the VERITAS collaboration.
The extension of the software for the reconstruction of MC events from CTA was only possible due to the support and contributions from the CTA MC team.
The following institutions and organisations supported the developers over the past 14 years: University of Leeds, Alexander-von-Humboldt Foundation, University of Delaware, National Science Foundation, McGill University, DESY, and the Helmholtz Association. 

These proceedings went through the review process of the VERITAS collaboration and the CTA consortium. 
We acknowledge the valuable input from the reviewers.

\end{document}